# 2D GaSe-Based Single-Pixel Spectrometer via Electro-Optical Barrier Co-Modulation

Shibesh Pramanik[1], Rishabh Sahoo[1], Arnab Mondal[2], Tithi Saha[2], Ankush Bag[2], Vibhav Bharadwaj Shivakumar[1], Rishi Maiti[1*]
[1]Department of Physics, Indian Institute of Technology, Guwahati, Assam, India, 781309
[2]Department of Electronics and electrical engineering, Indian Institute of Technology, Guwahati, Assam, India, 781309
*rmaiti@iitg.ac.in

## Abstract

Driven by the growing demand for miniaturized spectrometers for in-situ analysis, and point-of-care diagnostics, conventional spectrometers are often constrained by bulky architectures and pathlength-limited spectral resolution. Achieving high-resolution, single-pixel computational spectrometers is therefore critical for the realization of compact, on-chip systems. Here, we report a single-pixel spectrometer enabled by a single 2D material; few-layer GaSe-based photodetector, in which the Schottky barrier height modulation, governed jointly by applied bias and optical excitation, provides an efficient mechanism for spectral encoding without the need for bulky dispersive elements. The device exhibits a high peak-wavelength accuracy of ~0.78 nm across a broad operational bandwidth (300-700 nm) within a compact footprint of ~100 µm² and resolves closely spaced spectral features with separations down to ~5 nm. The device operates at low bias (±4V) with an ultralow dark current density ~0.3 pA/µm² at 4V bias. These results establish a simple, scalable route toward compact, cost-effective spectroscopic systems for on-chip spectral sensing and portable hyperspectral imaging applications.

## 1.| Introduction

Spectroscopy is a fundamental analytical technique widely used across diverse fields ranging from healthcare applications such as early diagnosis, and food quality assessment-to environmental monitoring [1,2]. Conventional spectrometers employed in these applications are often limited by their bulky size, excessive cost, and lack of portability since they need several dispersive optical elements (e.g. motorized gratings, mirrors, interferometers), and thousands of detectors or filter arrays. While these benchtop spectrometers deliver excellent spectral resolution and broad wavelength coverage [3], the lack of portability and integration capability prevents them from being widely utilized in modern in-situ portable systems such as consumer electronics, wearable devices, drones and remote sensors.

In recent years, various approaches have been introduced on miniaturized spectrometers to make it more affordable and compact by substituting bulky dispersive optical elements with a smaller number of tunable filter or detector arrays (spatial modulation) and leveraging various computational algorithms for the spectral reconstruction [1]. Examples of such approaches include single nanowire spectrometer with bandgap gradation [4], plasmonic nanoparticle in-cavity micro-filter array [5], structurally colored colloidal dot filter array [6], bandgap-engineered semiconductor spectrometers [7], folded metasurface spectrometer [8] and in-situ perovskite modulation-based spectrometer [9].

However, the limitation of the efficiency and practicality of these reported computational spectrometers can be attributed to three main factors. Firstly, the reliance of array-based architecture increases device footprint ($> 10\ \mathrm{cm}^2$) and fabrication complexity, while also introducing challenges in terms of pixel-to-pixel nonuniformity, alignment accuracy, and calibration stability. Secondly, their spectral resolution, operation bandwidth are also constrained by the number of integrated detectors or filters, bandgap modulation, and in certain cases the need for cryogenic operating conditions [10]. Lastly, operating voltage of most of the devices are relatively high due to the multi-gate configurations based spectral encoders [11]. Considering the aforementioned issues, a material platform capable of intrinsic and dynamic spectral tunability within a compact device footprint is highly desirable. In this regard, two-dimensional (2D) van-der Waals (vdW) materials have attracted significant attention as promising candidates owing to their atomically thin interfaces, strong light-matter coupling, broadband absorption capability, and integrability in CMOS platforms [12-14]. In addition, their layered dependent band structure and reduced dielectric screening enable dynamic tunability in the optoelectronic properties by external perturbation (like electric filed [15,16], strain [17,18]), which is very crucial for such reconfigurable spectrometry [4-9]. Importantly, these materials pave the way towards temporal modulation-based computational spectrometers within a single detector platform, where the spectral response of the device can be tuned dynamically using applied electrical bias, without the need for detector or filter arrays [19-24].

Our work therefore focuses on demonstrating a miniaturized spectrometer with a high signal-to-noise ratio based on a single photodetector, and on investigating the effect of light-induced dynamic barrier modulation at a metal-semiconductor (GaSe)-metal (MSM) junction. This approach provides insights into tunable photoresponse with a simple fabrication step and serves as an efficient control knob for spectral encoding, which has not been explored previously. In this work, we demonstrate a reconfigurable-type single pixel spectrometer over a broad operational spectral range (300-700 nm) using a layered GaSe based Schottky photodetector by strategically employing light induced barrier modulation that achieves high spectral reconstruction accuracy (~0.78 nm) within a minimal footprint (~100 $\mu m^2$). In contrast to the previously reported 2D vdW material-based spectrometers (e.g. pure BP [19], BP/$MoS_2$ [20] , $ReSe_2$/Au/$WSe_2$ [21], $ReSe_2$/$SnS_2$ [22], $NbTe_2$/InSe [23], $WSe_2$/$MoS_2$ [24]) that rely on the heterostructures composed of two different bandgap materials and typically require complex multi gate configurations with high gate voltages ($> 10$ V) to modulate the wavelength dependent spectral responses, our approach exploits a substantially simplified two-terminal device architecture for tunable Schottky barrier induced nonlinearly correlated photo-responses using low source-drain voltage range (-4 to 4V). Notably, the device exhibits ultralow dark current density (~0.3 pA/$\mu m^2$ at 4V bias) with high signal-to-noise ratio (~48 dB) of the spectrometer which facilitates its implementation in low-light level detection. These miniaturized spectrometers could provide a pathway towards lab-on-a-chip-, drone-, or satellite-based detection of subtle signals in applications such as material identification, biomedical sensing, and hyperspectral imaging.

## 2.| Results

### 2.1| Device Fabrication and Characterization

For the device fabrication few-layers GaSe is employed as an active material owing to its unique combination of features such as, direct band gap (~2 eV) induced strong light absorption in few-layers limit unlike transition metal dichalcogenides which shows direct bandgap in monolayer form [25,26], relatively low intrinsic carrier concentration (~$10^{14}$-$10^{16}$ $cm^{-3}$) [27], and moderate carrier mobility (~10-100 $cm^2V^{-1}s^{-1}$) [28], enabling low dark current (high specific detectivity) and high responsivity which is essential for spectral sensing and imaging applications. Schematic layout of our proposed device is shown in Figure 1a, consisting of hexagonal boron nitride (hBN)-encapsulated GaSe diode on a Si substrate with 300 nm $SiO_2$ layer on top. A bright field optical microscope image of the corresponding device is shown in Figure 1b. Detailed information of the device fabrication is described in the methodology section (4.1), and the step-by-step process is displayed in the supplementary Figure S1. The multilayer GaSe flakes were transferred directly on top of the prefabricated gold (Au) electrodes using our home-built dry transfer setup. This deterministic assembly strategy offers two distinct advantages over conventional fabrication: first, it circumvents the chemical contamination and unintentional doping typically introduced by various solvents used in lithography process. Second, it eliminates the lattice degradation associated with high-energy metal evaporation. By preserving the pristine form of the 2D lattice, this approach effectively suppresses Fermi-level pinning by minimizing defect-induced mid-gap states at the contact interface [29]. Atomic force microscopy (AFM) image in Figure 1c highlights the typical thickness of the GaSe flake as ~52 nm, confirmed from the measured height profile across the selected line. Figure 1d presents the room temperature PL spectra under 532 nm laser, where the encapsulated GaSe shows significantly higher intensity than bare GaSe, suggesting reduced defect-assisted non-radiative recombination [30,31]. The peak-wavelength (~620 nm) corresponds to the bandgap of the active material (GaSe), which is ~2 eV, consistent with the earlier reports [28,32]. Raman spectroscopy is also carried out to evaluate the crystalline quality of the exfoliated flake, presented in the supporting Figure S2. These results suggest that hBN encapsulation typically preserves the intrinsic optical properties of the material, which is crucial for achieving high photoresponsivity of the device. Scanning electron microscopy (SEM) image combined with energy-dispersive x-ray spectroscopy (EDS) mapping on a representative GaSe flake (Figure 1e) distinctly verifies the presence of Gallium (Ga) and Selenium (Se) with the ratios consistent with GaSe.

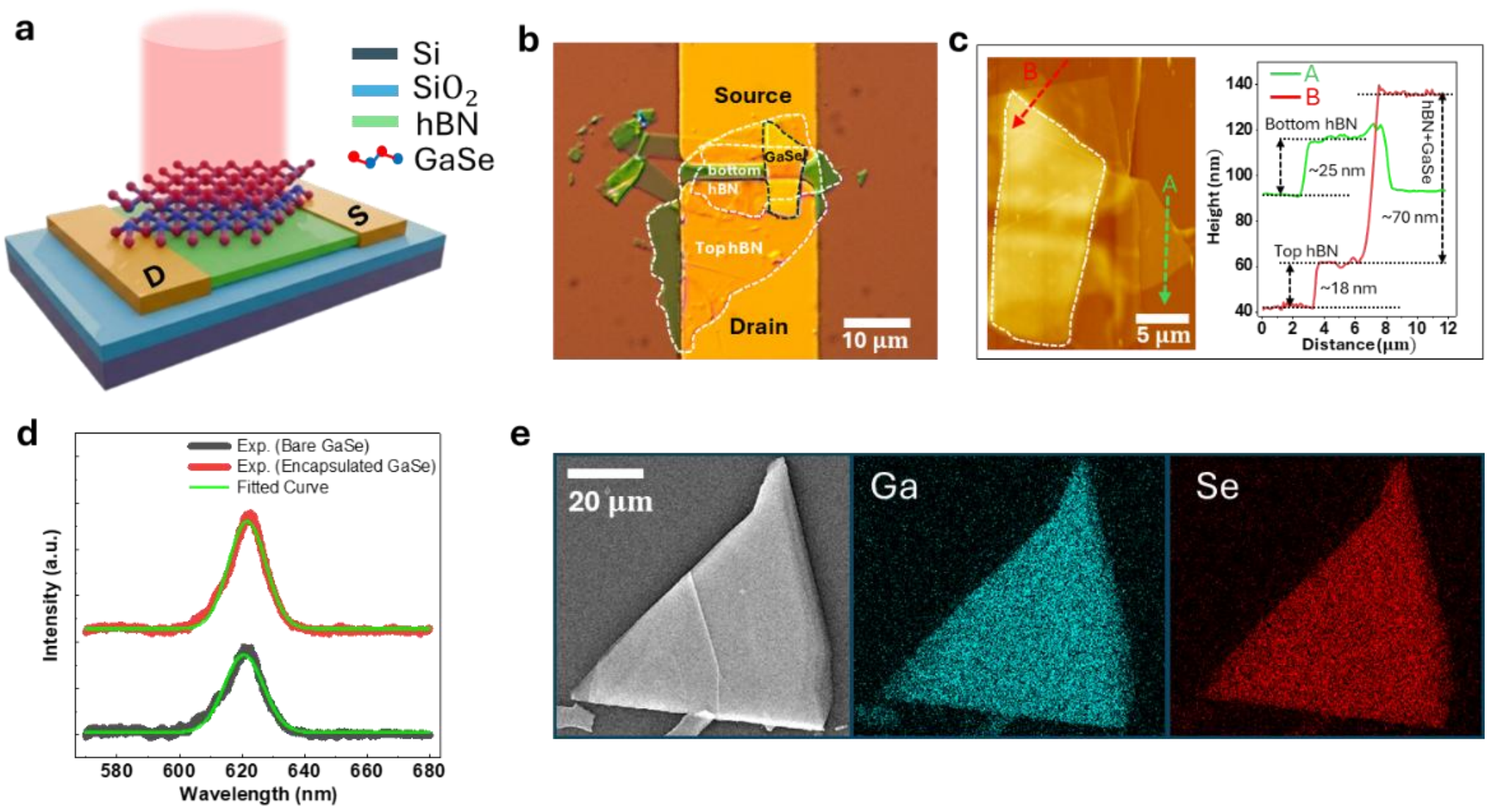

**Figure 1| Device architecture and its structural characterization.** **a**, Schematic of our metal (Au)-semiconductor (GaSe)-metal (Au) based diode with the hBN encapsulation. The top hBN is not shown in illustration. **b**, Corresponding bright field optical microscope image, where bottom hBN, top hBN and GaSe are highlighted with white and black dotted line, respectively. **c**, AFM image of the corresponding flakes with the extracted height profile varying along the green and red dotted line. The thickness of bottom hBN, top hBN and GaSe are ~25 nm, ~18 nm and ~52 nm, respectively. **d**, Photoluminescence spectrum of the corresponding material (GaSe), indicating the band gap ~2 eV. **e**, Scanning electron microscopy (SEM) image combined with the energy-dispersive x-ray spectroscopy (EDS) mapping on a representative GaSe flake.

We measure the I-V characteristics of our fabricated Au/GaSe/Au device in the voltage range from -4 to 4V (Figure 2a). The device exhibits rectifying behavior with a dark to photocurrent ratio of $\sim 10^4$ (inset of Figure 2a) due to extremely low dark current (~30 pA at 4V), thus enabling enhance device detectivity (D*$\sim 10^{11}$ Jones) by reducing the noise floor. Figure 2a illustrates the current-voltage (I-V) characteristic of the device, showing a symmetric I-V curve with respect to $V_{ds}$, which suggest that the device is governed by two oppositely oriented Schottky junction at the metal-semiconductor (GaSe/Au) interface. Gold (Au) contacts are selected due to their high work function (~5.1-5.5 eV) [33] and chemical stability, which favor Schottky barrier formation with GaSe having an electron affinity of ~3.6-4.2 eV [27,28]. While the Schottky-Mott rule predicts a barrier height $\varphi_B$~0.9 eV for typical metal-semiconductor interfaces, 3D metal-2D semiconductor junctions often exhibit significant deviations due to strong Fermi-level pinning [34]. In practice, Schottky barrier heights for various 2D materials have been shown to fall consistently below theoretical predictions, often residing in the range of 0-0.2 eV [35,36]. By employing a transfer-based assembly of 2D materials onto prefabricated metal contacts, our approach suppresses Fermi-level pinning and preserves the intrinsic electronic properties of the interface [29,36].

The contact of Au/GaSe enables fermi level alignment at equilibrium (Figure 2b (i)) and the formation of Schottky barrier height ($\varphi_B$~0.67 eV), which effectively suppresses the carrier injection and charge transport across the GaSe-channel, resulting in low dark current (~0.1 pA at 0V) near zero-bias conditions. Integrating high work-function metal electrodes (Au) with an n-doped 2D semiconductor results in a large Schottky barrier for majority carriers (electrons) and a negligible barrier for photoexcited minority carriers (holes) [37,38]. As a result, the elevated barrier suppresses dark current while the photoexcited holes are efficiently collected, leading to enhanced photocurrent. When source-drain bias is applied across the metal-semiconductor junction, electrostatic potential becomes asymmetric, causing a reduction of effective Schottky barrier height (Figure 2b (ii)) which enables directional carrier transport across the device and resulting in a measurable dark current. Upon illuminating light on the device (Figure 2b (iii)), photon-generated carriers accumulate near the interfaces, screening the built-in electric field and narrowing the depletion region, leading to a further reduction in the effective Schottky barrier height ($\varphi'_B$). The combined effect of applied bias and light illumination enhances carrier injection and separation, producing a measurable photocurrent across the device. The wavelength-induced transport behavior is experimentally underlined in Figure 2c, where the photocurrents of the device are recorded at a fixed bias voltage (+4V) over a broad spectrum range from 300 to 700 nm. The variations in photocurrent can be attributed to different excitation wavelengths induce distinct transport responses even at a constant applied bias. The origin of this variation can be explained from the intrinsic absorption characteristics of few layers GaSe [39,40] and the corresponding excess photo-carrier generation and recombination. It is worth noting that a sharp peak appears ~620 nm (Figure 2c), corresponding to the photon energy of ~2 eV, which is close to the typical bandgap of few layers GaSe also confirmed from light illumination, showing the pronounced photo-induced current enhancement (~$10^4$). **b**, Schematic energy band diagram of the device, illustrating the Schottky barrier formation ($\varphi_B$) at equilibrium state (i) and its modulation ($\varphi'_B$) under bias and light illumination (ii and iii), where photo-generated charge transfer and band bending lead to effective reduction in barrier height. $E_F$ corresponds to fermi energy of the metal. **c**, Wavelength dependent photocurrent of the device measured across the operational spectral range of 300-700 nm at a fixed bias of 4V. The inset showing the variations in photocurrent clearly near the band gap of the material (~2 eV). **d**, Two-dimensional (2D) heat map of the effective Schottky barrier modulation ($\Delta\varphi_B$) as a function of applied bias ($V_{ds}$) and excitation wavelength (λ). **e**, extracted barrier height ($\Delta\varphi_B$) profile as a function of applied bias (-4V to +4V) for two representative wavelengths (350 nm & 620 nm), revealing voltage tunable barrier modulation. Corresponding wavelengths are indicated by red dotted line in the 2D color plot. **f**, Excitation wavelength dependent effective barrier height ($\Delta\varphi_B$) profile at two different applied biases ($-2V, 3V$), indicating photon energy dependent carrier transport. Corresponding biases are highlighted by white dotted line.

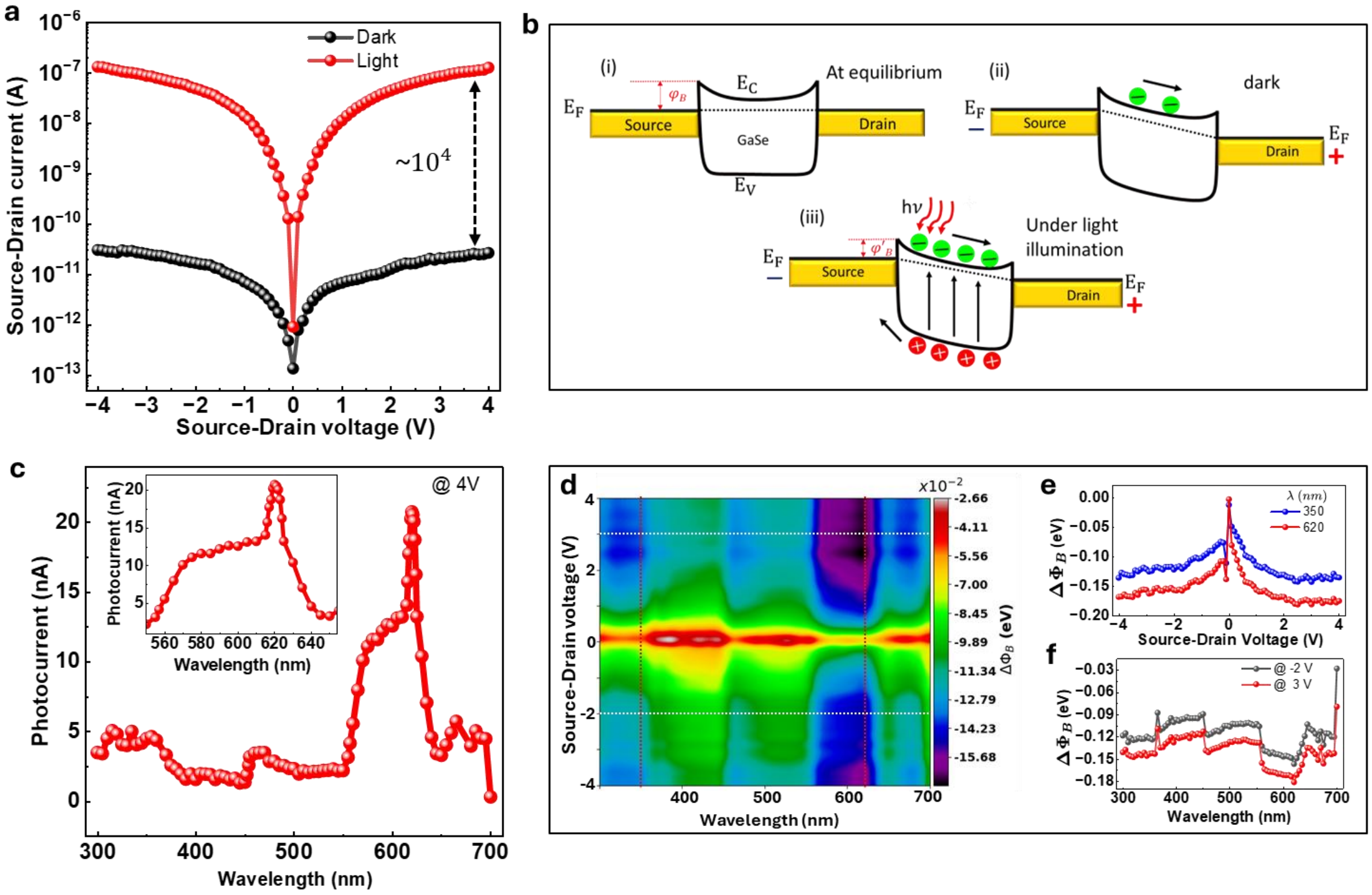

**Figure 2| Optoelectronic characterization and Schottky barrier modulation of the MSM photodetector**. **a**, Current-voltage (I-V) characteristics (semi-log plot) of the Au/GaSe/Au device under dark condition and white

origin of this variation can be explained from the intrinsic absorption characteristics of few layers GaSe [39,40] and the corresponding excess photo-carrier generation and recombination. It is worth noting that a sharp peak appears ~620 nm (Figure 2c), corresponding to the photon energy of ~2 eV, which is close to the typical bandgap of few layer GaSe also confirmed from the PL spectra shown in Figure 1f. In this region the absorption coefficient remains sufficiently high, leading to strong inter-band absorption and highest photoresponse. However, the sharp decline of photocurrent beyond 620 nm can be attributed as the GaSe absorption decreases significantly due to weak interaction of sub bandgap photons. For completeness, the current-voltage (I-V) characteristics measured across the full voltage range (-4 to +4V) for different illuminated wavelengths are presented in the supporting information (supplementary Figure S3), suggest that the photon energy modifies the interfacial barrier in a wavelength-dependent manner rather than producing a uniform photoconductive gain, resulting in distinct and nonlinearly-correlated photoresponses. Further, to evaluate the photodetection performance of the device across the entire wavelength range (300-700 nm), specific detectivity is calculated in the supporting figure S3.1. In addition, linear dynamic range (LDR) and transient response characteristics (including rise and fall times) of the device are also analyzed systematically and presented in supporting figure S3.2, indicating strong potential of the device to be used as a single-pixel spectrometer.

We further investigated the photocurrent modulation mechanism to evaluate the light induced effective Schottky barrier height modulation ($\Delta\varphi_B$) within a thermionic emission framework (supplementary text S4, Figure S4) by using the dark and illumination current responses in the following equation [41,42,43].

$$\Delta\varphi_{Bi} = \frac{K_B T}{q} \ln\left\{\frac{I_{dark}(V_i)}{I_{light}(V_i,\lambda_i)}\right\} \quad (1)$$

Where, $\Delta\varphi_{Bi}$ term represents the modulation of Schottky barrier height under the illumination of light, q stands for elemental charge, $I_{Dark}$ and $I_{Light}$ are the dark and light current respectively. The extracted $\Delta\varphi_{Bi}$ ($V_i, \lambda_i$) is clearly reflected in the two-dimensional contour map (Figure 2d), provides a deep insight into the coupled electrical and optical modulation. For better understanding representative voltage and wavelength dependent profiles are extracted from the contour map, as illustrated in Figure 2e and 2f, respectively. As shown in Figure 2e, the systematic variation of the effective barrier height with increasing bias voltage for both the wavelengths (350 and 620 nm), consistent with electric field assisted barrier thinning and enhanced carrier injection across the MSM junction. Similarly, the extracted profiles for two different biases (-2V and 3V) is presented in Figure 2f, where the variations in effective barrier height exhibits a pronounced wavelength dependency with the most significant barrier reduction occurs at ~620 nm (corresponds to the band gap of the material ~2 eV), follows the experimentally obtained photocurrent spectrum (Figure 2c). Therefore, the photon energy influences the spatial distribution and dynamics of photo generated carriers within the GaSe layer, leading to wavelength-selective modifications of band-bending at the metal-semiconductor interface, highlighting that the barrier height is not only a static interfacial parameter, but rather a dynamically tunable quantity governed by the interplay of both the applied electrical bias and the photon energy. Collectively, these results suggest that the Schottky contact function as an actively engineered transport filter whose effective barrier encodes spectral information. Hence, this electro-optically co-tuned barrier modulation constitutes the physical basis for generating sufficient diversity in the wavelength-selective photo current response matrix, which form the foundation for spectral encoding and enable computational reconstruction of unknown incident light in a single material architecture instead of relying on heterojunction engineering or external gating, which is explored in detail in the following section.

## 2.2| Spectrometer demonstration with the operational principle

Figure 3a depicts the operational mechanism of our spectrometer, comprising two sequential steps: a. first the spectral encoding and b. decoding via spectral reconstruction. By applying a variable external bias between the source and drain electrodes, the optoelectronic responses of a two terminal device can be modulated systematically. In the encoding step, as illustrated in Figure 3a, the device is characterized initially under a series of known monochromatic light inputs (known wavelengths), resulting in a unique photoresponsivity matrix R for the device, expressed as a function of both the input wavelengths (λ) and source-drain voltages ($V_{ds}$), with its electrically tunable elements ($R_{ij}$) is defined as:

$$R_{ij} = \frac{I_{ph}(\lambda_i, V_{dsj})}{P_{eff}} \qquad (2)$$

Here $P_{eff}$ is the effective optical power of the input spectrum (for detailed calculation see the supplementary Figure S5) and $I_{ph}$ is the photocurrent of the device, defined as $I_{ph} = I_{light} - I_{dark}$ ,with $I_{light}$ and $I_{dark}$ representing the source-drain current with and without light illumination at a fixed bias voltage. In the decoding step, the electrical response (photocurrent) of an unknown light is used in an algorithm along with the predefined response matrix to achieve spectral reconstruction.

For the operation of our spectrometer, here we have demonstrated first the narrow band spectrum reconstruction using monochromatic light as an input. In the learning stage, first we measured the photoresponses of the device by varying the incident excitation wavelength from 300 to 700 nm in a step of 5 nm and then calculated the responsivity (R) as a function of both the spectrum wavelengths and applied source-drain voltages ($V_{ds}$), shown in Figure 3b with the color contour map, exhibit a rich and distinct structure governed by the joint effect of excitation wavelength and applied bias in terms of photo-carrier generation and Schottky barrier transport in the Au/GaSe/Au device. Notably, the observed variation in the responsivity matrix closely follows the contour map of Schottky barrier height modulation (presented before in Figure 2d), indicating that the spectral encoding directly originates from the dynamically tunable interfacial barrier. Once the spectral response matrix is encoded during the learning stage, our single device spectrometer can be employed to determine unknown incident spectra, as outlined in the above workflow diagram (Figure 3a). The bias-dependent photocurrent is recorded under the illumination of unknown input spectrum, and the spectral fingerprint is retrieved through a constrained least-square solution for spectral reconstruction using adaptive Tikhonov regularization algorithm [19-24] (complete algorithm flowchart is provided in supplementary Figure S6). Detailed comprehensive information's about the optical set up, as well as optoelectronic characterization are provided in the methodology section (4.3), and the measurement set up is illustrated schematically in the supporting information (Figure S7). Figure 3**c** illustrates the reconstruction of quasi-monochromatic spectra at 6 different wavelengths (370 nm, 430 nm, 500 nm, 550 nm, 600 nm, and 650 nm, respectively), alongside the reference spectra taken from a commercial spectrometer (GA11315, Firefly 4000: 350-1000 nm) as ground truth, enabling a direct comparison of the spectral accuracy and performance of our device. The reconstructed peak positions are reliably recovered across the entire tested spectral range (300 to 700 nm), having sufficient resolution to distinguish neighboring peaks down to a separation of 5 nm (Figure 3d). The average peak difference between the reconstructed and reference spectrum is about ~0.78 nm, shown in Figure 3e, indicating excellent reconstruction accuracy. The proposed spectrometer is further tested using two different LEDs as input light sources, shown in Figure 3f. The left panel displays the schematic of experimental set up, where the light emitting from the LED source is divided into two equal intensity components using a 50:50 beam splitter. One beam is directed towards the commercial spectrometer to obtain the reference spectra, while the other beam is coupled into the proposed spectrometer for spectral reconstruction and performance comparison. In Figure 3f (right panel), the reconstructed spectrum is compared with the corresponding reference spectra for the yellow, and red LEDs having central wavelengths of 592 and 625 nm, respectively. The reconstructed and reference spectra are in good agreement with respect to both the spectral peak positions and the full width at half maximum (FWHM), yielding root-mean-square deviation (RMSD) of ~0.83 nm and ~2.65 nm for the yellow and red LEDs, respectively. The formulation of RMSD is shown in the supporting information (Note 1).

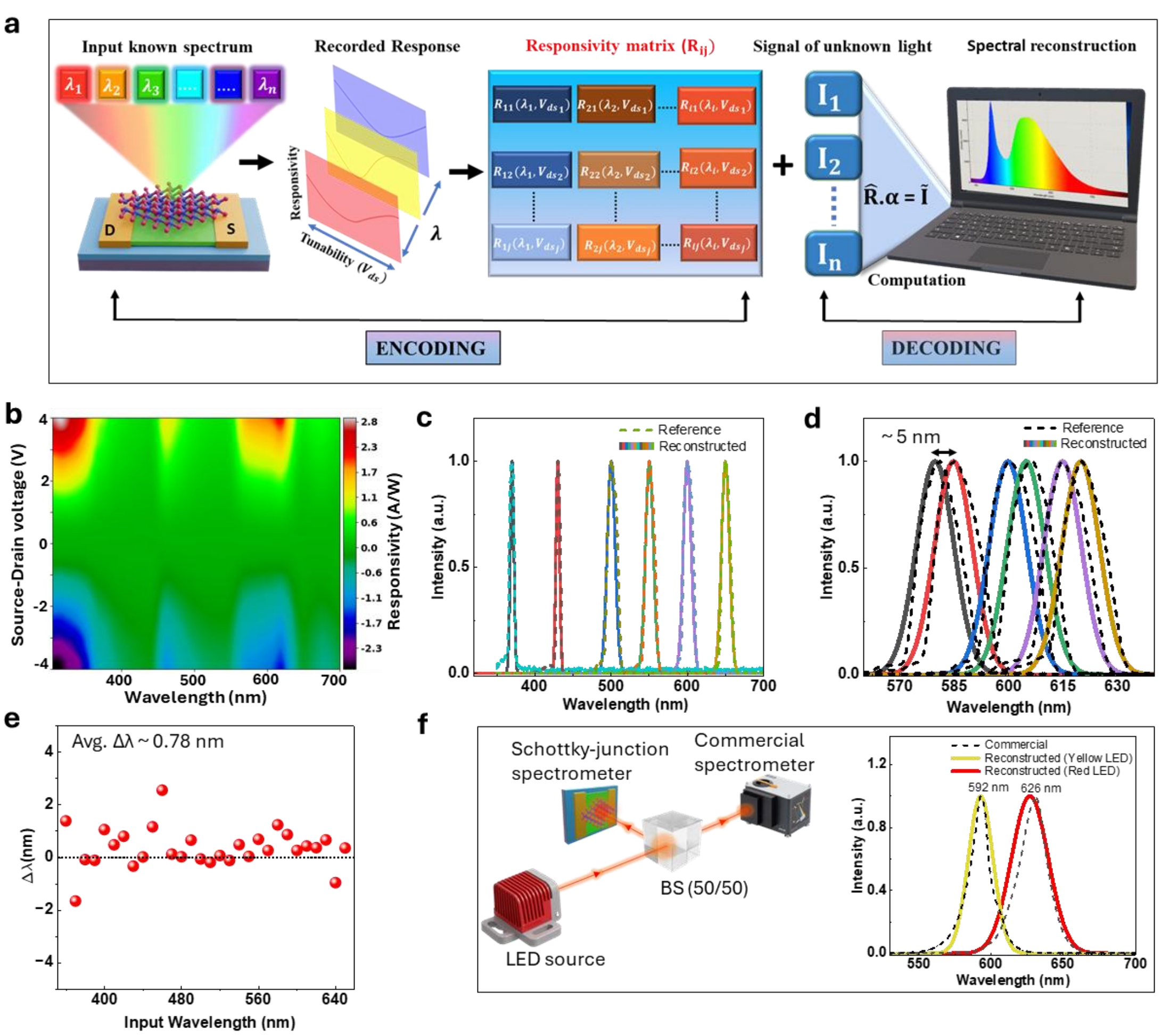

**Figure 3| Spectrometer demonstration along with the operational principle**. **a**, Schematic representation of the working principle of our spectrometer, highlighting the wavelength and voltage dependent ($V_{ds}$) spectral encoding (in terms of responsivity matrix $R_{ij}$) process and the subsequent decoding procedure for spectrum reconstruction. **b**, two-dimensional (2D) color contour map of experimentally obtained response matrix (81 different spectrum and 81 voltage values per spectrum). **c**, six quasi monochromatic spectra across the entire operational spectral range from 300 to 700 nm, reconstructed using our spectrometer (solid curves) and measured via a commercial spectrometer (dashed curves). **d**, reconstruction of closely spaced input wavelengths separated

by ~5 nm, Illustrating the spectral resolution limit of the device. **e**, the peak wavelength difference, or the reconstruction error (Δλ) between the reconstructed and reference spectra, with the average of Δλ indicating overall spectral reconstruction precision. **f**, **Left panel** displays the schematic illustration of the experimental configuration for simultaneous spectral acquisition of LED sources from a commercial spectrometer and the proposed device. **Right panel** reflects the reconstruction of two narrowband LED sources corresponding to yellow (~592 nm) and red (~625 nm) emission.

Additionally, broadband spectral reconstruction is carried out (supplementary Figure S7) to validate the capability of the proposed single-pixel spectrometer to recover complex spectral profile. The reconstructed spectrum closely overlapped with the reference spectrum taken from a commercial spectrometer. To further access the reproducibility of the spectrometer performances, device-to-device variations are investigated (supplementary S9; Figure S9.1 to S9.3) across multiple fabricated devices (Au/GaSe/Au) under the same circumstances.

## 2.3| Performance analysis (noise tolerance, stability, and repeatability) of the spectrometer

In order to evaluate the robustness of our spectrometer, we have further investigated systematically the effect of signal-to-noise ratio (SNR) on spectral reconstruction. Specifically, at lower signal-to-noise ratios (SNRs) the measured signals may exhibit significant fluctuations instead of expected smoothed profile, leading to inaccuracies in the reconstructed spectrum compared to the true values.

As shown in Figure 4a, at low SNR (~20 dB) (for more details follow Supplementary note2) the peak wavelength shift is almost 45 nm with respect to the ground truth, while for SNR greater than 40 dB the noise contribution is reduced, leading to more accurate spectral reconstruction (i.e., the spectrums closely overlapped each other).

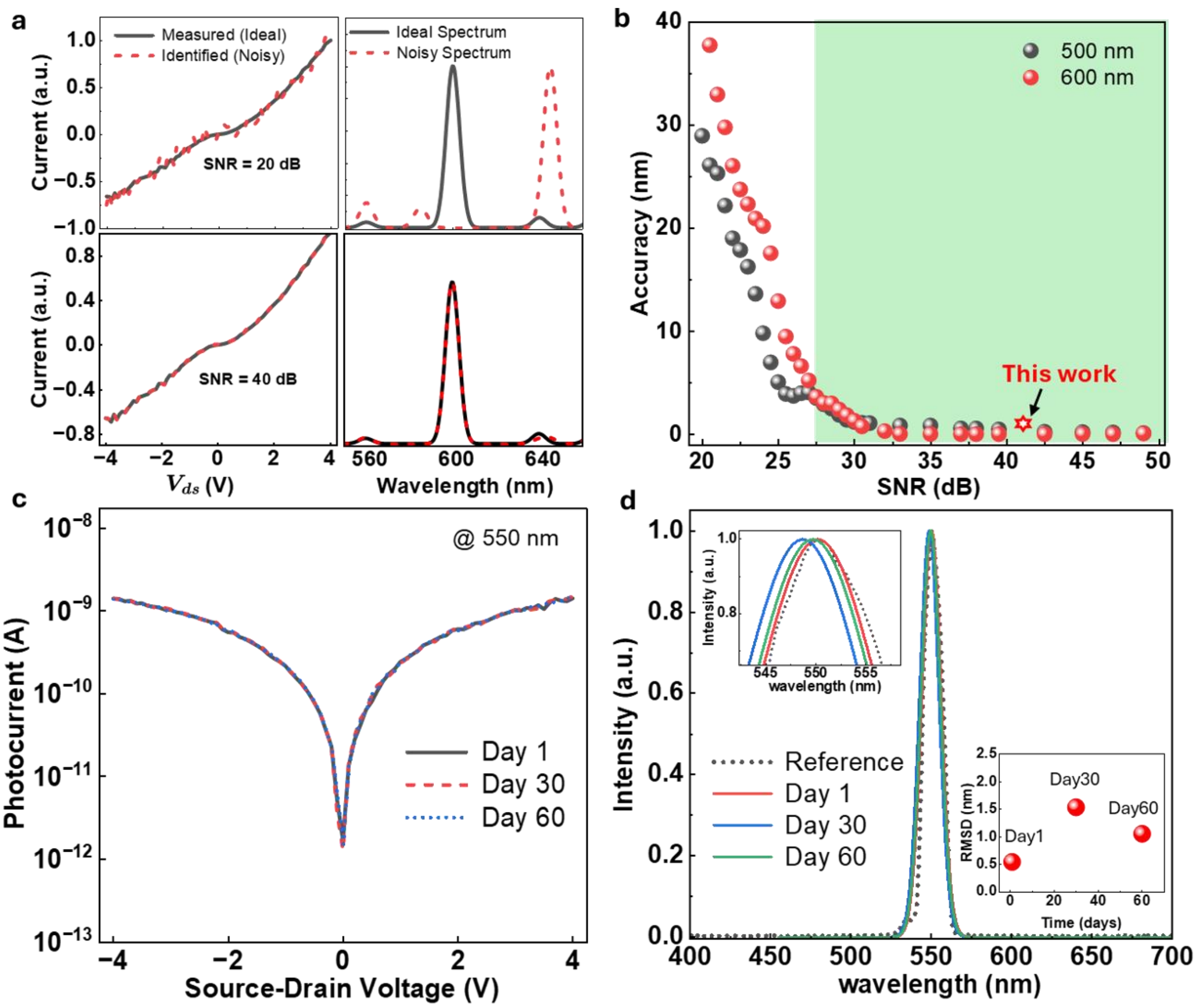

**Figure 4| Noise tolerance along with the operational stability and repeatability of our spectrometer performance**. **a**, Impact of noise level on the measured electrical signal for different SNRs (20 dB & 40 dB). **b**, Reconstruction accuracy as a function of SNR for two representative wavelengths (500 nm & 600 nm). The red star highlights the performance of our device, and the shaded region indicates experimentally relevant regime for the operation of our spectrometer. **c**, Repeating measurement of current-voltage (I-V) characteristics of our device under identical light sources (550 nm) from day 1, day 30 and day 60, respectively. **d**, reconstructed spectra of the same source of light irrespective of the mentioned period. The inset showing distinct peak wavelengths (top left corner) at 550.2 nm, 548.7 nm and 549.5 nm, and the RMSD (bottom-right corner) of 0.54 nm, 1.53 nm and 1.05 nm, respectively.

As depicted in Figure 4b, the simulated identification accuracy shows a clear dependence on the SNR for two different wavelengths (500 and 600 nm). It is observed that above 27 dB of noise level (Shadow area), the accuracy of monochromatic light identification is almost equal to its peak wavelength's separation limit ~5 nm (Figure 3d), which can be defined as a high-precision identification. Our experimental results are calculated (marked as red star) and lies in the range from ~ 48 to ~ 50 dB, with the average reconstruction accuracy ~ 0.78 nm.

In addition to the noise tolerance, the operational stability and repeatability of our spectrometer are realized systematically by evaluating both the long-term and short-term measurements. To evaluate the long-term operational stability, the device photoresponses are monitored over an extended period around two months. Instead of repeating the complete response matrix measurements, the photocurrents are recorded at the selected wavelengths to examine variations in the device.

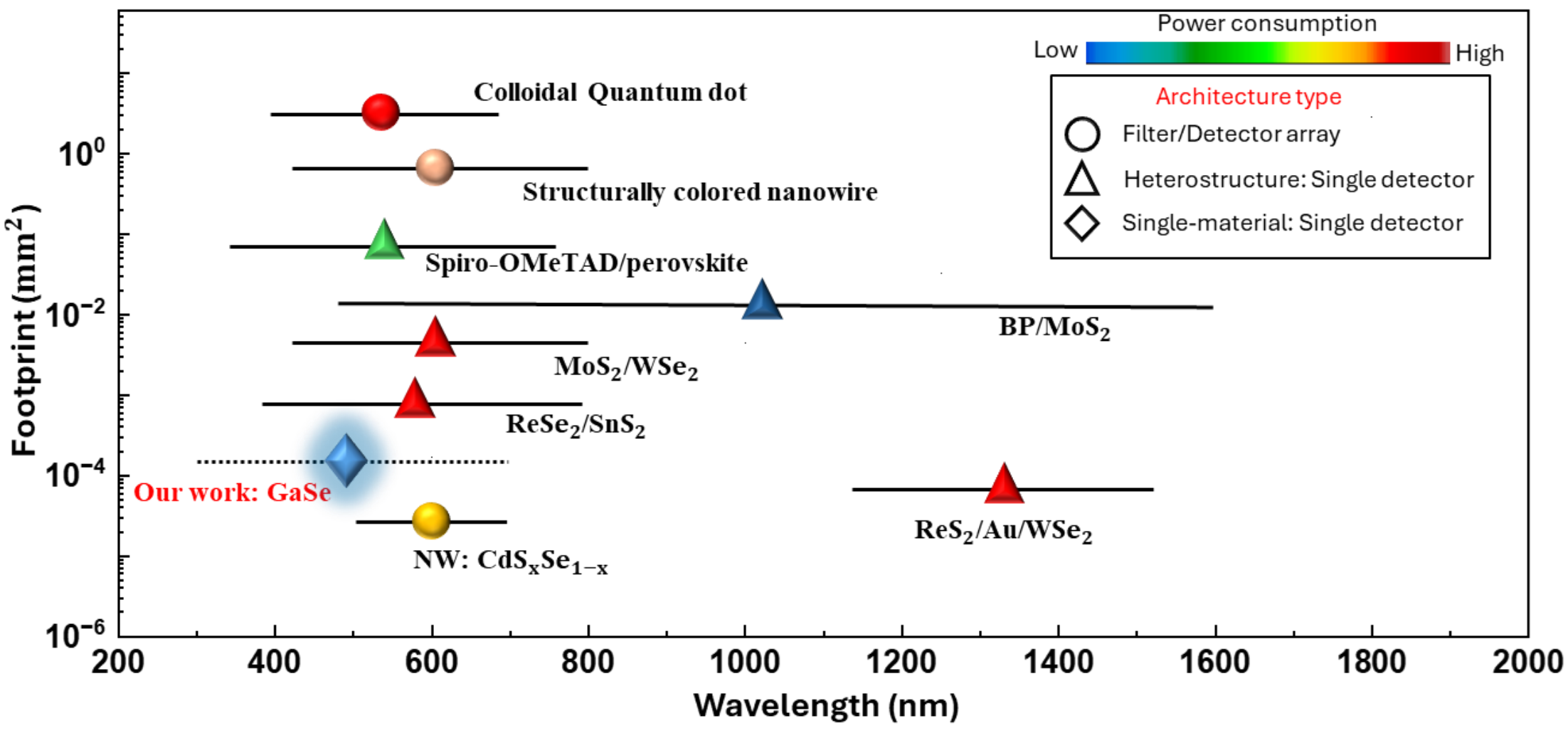

**Figure 5| Performance comparison** of our spectrometer with the other reported literatures in terms of operational spectral range, device footprint, and relative power consumption. $MoS_2/WSe_2$ vdWH (van-der Waals heterostructure) [24], $ReSe_2/SnS_2$ vdWH [22], Graded nanowire $CdS_xSe_{1-x}$ spectrometer [4], $BP/MoS_2$ vdWH [20], spiro-OMeTAD/perovskite [44], structurally colored nanowire [45], quantum dots spectrometer [6], and $ReS_2/Au/WSe_2$ vdWH [21]. The horizontal solid lines represent the operational spectral range. Three different geometric symbols are used to denote the architecture used in the corresponding literatures and their inside solid colours indicates typical power consumption of the device.

performance over the time. Figure 4c presents the measured photocurrent of the device at a representative wavelength of 550 nm on day 1, day 30 and day 60 respectively, under identical experimental conditions. The nearly identical responses suggest that the device maintain high stability over the entire measurement period. Further, to verify the reliability of spectral reconstruction over the same period, the incident spectrum at 550 nm is reconstructed (Figure 4d) using the response data of Figure 4c. The reconstructed spectra closely overlap with the reference spectra having the RMSD of 0.54 nm, 1.53 nm and 1.05 nm respectively (inset of Figure 4d, bottom-right corner). Along with the long-term stability, the short-term repeatability of the device is also investigated through 100 consecutive current-voltage measurements under both the dark and illuminated conditions, respectively (supporting Figure S10). Overall, these results highlight the robustness of our spectrometer for spectral sensing applications.

To place the performance of our spectrometer in context, we benchmarked our work against the previously reported miniaturized spectrometers in Figure 5, based on key performance metrics, including spectral range, device footprint and relative power consumption by the device. Notably, the proposed spectrometer demonstrated high peak wavelength accuracy over a broad operational window spanning from UV to visible regions while maintaining compact device footprint and low power consumption. Unlike existing approaches that rely on complex heterostructures and high gate voltage tuning, our spectrometer employs a single-material architecture with low source-drain bias modulation, enabling its scalability and practical implementation.

### 2.4| Spectral imaging demonstration

To further validate the applicability of the proposed single-pixel spectrometer, a proof-of-concept 2D hyperspectral imaging demonstration is conducted using the experimentally obtained responsivity matrix R ($\lambda$, $V_{ds}$). Owing to the absence of a special physical scanning stage in our lab, the imaging process is implemented through simulation, where the measured device responses are employed to reconstruct spatially resolved spectral information. The detailed two stage reconstruction processes are outlined in the supporting information (supplementary Note 3).

Figure 6a, displays the schematic of the simulated spatial scanning process, where the responses of the device are sequentially acquired across a virtual scene using a 150 × 150-pixel spatial grid, constructed with a logo as the target object. The logo consists of multiple color regions primarily blue, yellow, and red, along with white circular features, forming a spatially

varying spectral distribution (Figure 6b). These spatial regions are assigned distinct yet partially overlapping spectral transmission profiles, denoted as filter 1 (blue; 390 to 500 nm), filter 2 (yellow; 510 to 590 nm), and filter 3 (red; 580 to 680 nm), respectively (Fig. 6**b**), to examine the spectral discrimination capability of the spectrometer. Utilizing the measured responsivity matrix as the key encoding element of the system, image reconstructions are performed across the wavelength range of 380 to 680 nm. In Figure 6c (top), the reconstructed 2D hyperspectral images are presented at different wavelengths, where the spatial distribution corresponding to the different colored regions are clearly resolved, and most importantly these images preserve the relative intensity variations, demonstrating accurate recovery of both spectral and amplitude information. Further, to evaluate the reconstruction fidelity, corresponding ground truth images are also displayed in Figure 6c (bottom). For the quantitative evaluation of the image quality, brightness differences are introduced between the reconstructed and ground truth images and defined as $\Delta \mathrm{I} = \mathrm{I_{Rec}} - \mathrm{I_{Gt}}$, where $\mathrm{I_{Rec}}$ and $\mathrm{I_{Gt}}$ are the normalized intensity of the reconstructed image and the corresponding ground truth image, respectively. The strong agreement between the reconstructed and true spatial patterns along with the preserved intensity contrast (minimum deviations) highlights the robustness of our single-pixel spectrometer for image reconstruction.

Overall, this proof-of-concept 2D hyper-spectral imaging demonstration underscores the potential of the proposed single-pixel spectrometer to perform compact and high sensitivity spectral imaging without relying on bulky dispersive optical components or multi-pixel detectors, paving the way to be utilized in miniaturized and compact hyper-spectral sensing platforms.

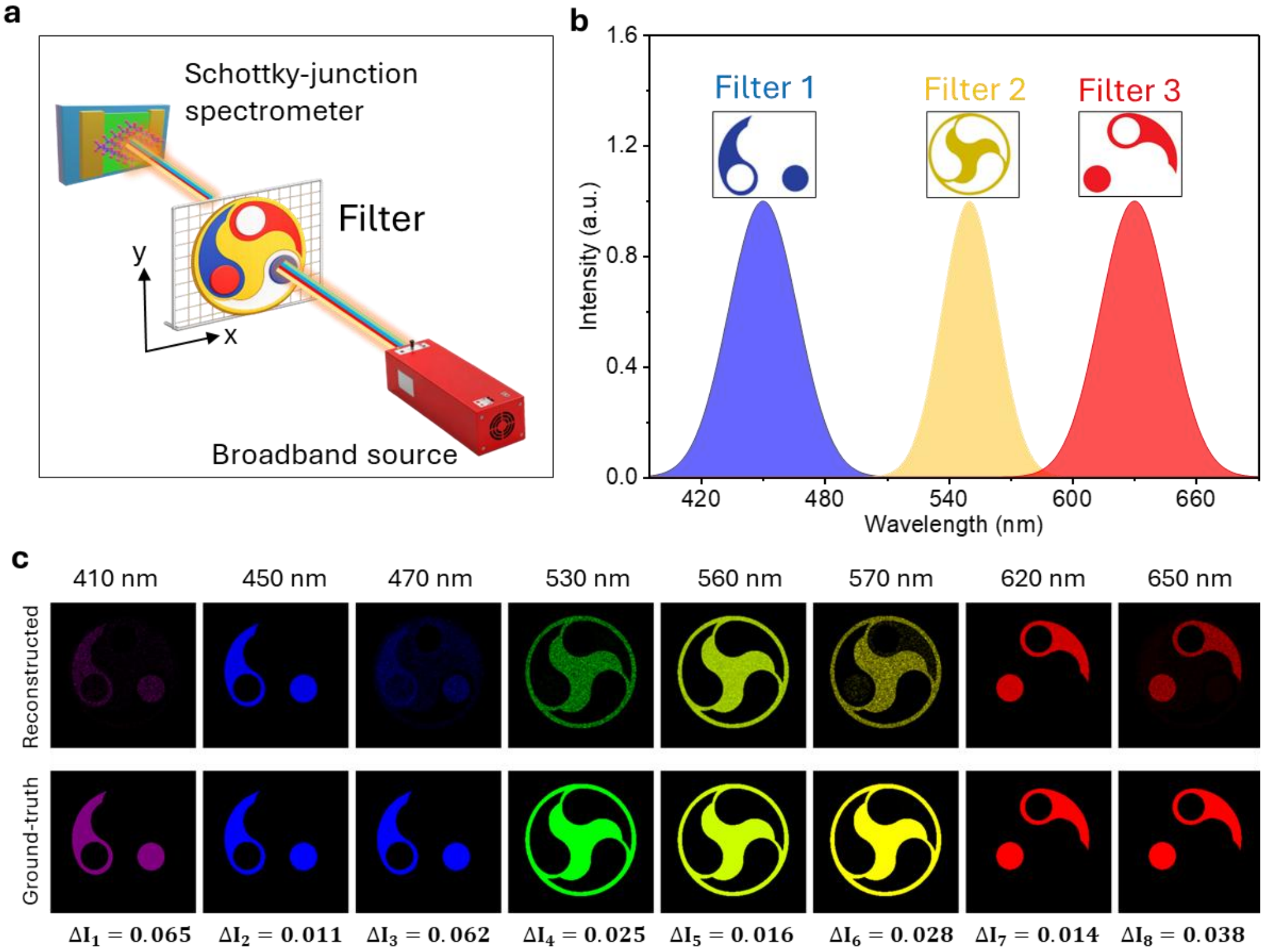

**Figure 6| 2D Hyperspectral imaging. a**, Configuration of hyperspectral imaging process using our single-pixel spectrometer with a spatial scanning method. A broadband light source filtered with a color image (logo) and exposed to our device for spectral imaging. **b**, Reconstructed transmission spectra corresponding to the color filters, denoted as Filter 1 (from 390 to 500 nm) and filter 2 (from 510 to 590 nm), and filter3 (580 to 680 nm), respectively. **c**, Reconstructed hyperspectral images across the selected wavelengths, compared with the corresponding ground truth values in terms brightness difference (ΔI).

## 3.| Discussion

In summary, here we have successfully demonstrated a reconfigurable-type single pixel spectrometer based on a single semiconductor (GaSe) device architecture, achieving a broad operational bandwidth within an ultra-compact device footprint. The spectrometer is electrically reconfigured by exploiting the applied voltage and wavelength dependent photoresponses, instead of using any filter or array of detectors or other dispersive optical components to achieve high resolution with sub-nanometers accuracy. Furthermore, the low operating voltage together with ultracompact device footprint makes this platform highly suitable for the integration into compact and energy-efficient devices. Future work will explore the engineering of metal-semiconductor interface and contact geometry to enhance the Schottky barrier tunability for improved spectral resolution, extended operational bandwidth and to make the device suitable for ultralow light detection with high reconstruction fidelity. These attributes pave the way for its usability in emerging consumer photonics and affordable on-chip spectral imaging systems, with broad applicability in areas such as space missions (where chip-scale spectral sensing can play a crucial role in in-situ material identification and analysis under resource-constrained conditions), remote sensing, and environmental monitoring.

## 4.| Methodology:

### 4.1| Material exfoliation and device fabrication

For the device fabrication we use Si substrate having 300 nm $SiO_2$ layer on top. Prior to electrode fabrication, the substrates are cleaned sequentially in acetone, isopropyl alcohol (IPA), and de-ionized water, followed by nitrogen blow drying. First few layers of GaSe and hBN flakes are mechanically exfoliated from bulk crystal (commercially purchased from HQ Graphene) using scotch tape (PDMS) and then transferred sequentially to the prefabricated electrodes on $Si/SiO_2$ substrate using our inhouse-built dry transfer set up. 60 nm thick source and drain electrodes are fabricated using standard photolithography system followed by thermal evaporation, first the positive photoresist (S1813) is spin coated (3000 rpm) on a cleaned $Si/SiO_2$ substrate and then patterned with photolithography (375 nm laser source). Afterwards, the sample is loaded in a high vacuum chamber of thermal evaporator for gold deposition, and finally by acetone lift off the electrodes are prepared.

### 4.2| Raman, Photoluminescence, and atomic force microscopy (AFM) measurement:

All the active vibrational fingerprints of the material are recorded at room temperature using a micro-Raman system (Make: Horiba Jobin Vyon, Model LabRam HR) followed by a 514 nm diode laser (He-Ne laser) excitation. The laser power is maintained as low as possible (~1 mW) to avoid any possible damage in the sample. The excitation laser is focused onto the sample surface through a high numerical aperture (~0.75) objective lens, and the scattered signal is also collected through the same. Further, the scattered signals are detected using charge-coupled-device (CCD) spectrometer under ambient conditions.

Atomic-force-microscopy (AFM) (Make: Oxford, Model: Cypher) measurements are carried out under tapping mode to determine the surface morphology and the thickness of the material. The height profiles are extracted from the AFM image using Gwyddion software.

### 4.3| Electrical and opto-electrical measurements:

Electrical measurements are carried out using a probe station equipped with a high precision source meter (Keithley 2450) for biasing and current read out. The device is placed under the microscope and probed electrically using two micro manipulators to ensure precise contact with the respective source-drain electrodes. All the measurements are performed under ambient conditions at room temperature. For opto-electrical measurements a broadband halogen light source coupled with a monochromator (use specs) is employed to generate monochromatic light of wavelength spanning from UV-Vis range (300-700 nm) with the spectral band width of 5 nm. The monochromator output spectrums are calibrated using a commercial spectrometer (our spec specification) to verify its peak wavelength accuracy and the full width half maxima (FWHM) as well.

**Acknowledgement**

The authors acknowledge financial support from the Science and Engineering Research Board (SERB) Department of Science and Technology (DST), Government of India, under the core Research Grant (CRG/2021/006815). The authors further acknowledge the central instrument facility (CIF) and the center for nanotechnology, Indian Institute of Technology Guwahati, for access to characterization and nanofabrication facilities.

## Authors contributions

RM conceived the idea, supervised the overall research and contributed to the final version of the manuscript. SP designed the experiments, fabricated the spectrometer device, carried out all the electro-optical measurements, and prepared the manuscript. RS developed the spectral reconstruction algorithm, and computational framework. AM and TS assisted with the device characterization. AB and VBS contributed to the final version of the manuscript.